\documentclass[a4paper, conference]{IEEEtran}
\IEEEoverridecommandlockouts
\usepackage{cite}
\usepackage{amsmath,amssymb,amsfonts}
\usepackage{algorithmic}
\usepackage{graphicx}
\usepackage{textcomp}
\usepackage{comment}
\usepackage{booktabs}
\usepackage{tabularx}
\usepackage{array}
\usepackage{xcolor}
\def\BibTeX{{\rm B\kern-.05em{\sc i\kern-.025em b}\kern-.08em
T\kern-.1667em\lower.7ex\hbox{E}\kern-.125emX}}

\usepackage{tikz}

\newcommand\copyrighttext{%
\footnotesize \textcopyright\ 2026 IEEE. Personal use of this material is permitted. Permission from IEEE must be obtained for all other uses, in any current or future media, including reprinting/republishing this material for advertising or promotional purposes, creating new collective works, for resale or redistribution to servers or lists, or reuse of any copyrighted component of this work in other works.}

\newcommand\copyrightnotice{%
    \begin{tikzpicture}[remember picture,overlay]
        \node[anchor=south,yshift=10pt] at (current page.south)
        {\fbox{\parbox{\dimexpr0.9\textwidth-\fboxsep-\fboxrule\relax}{\copyrighttext}}};
    \end{tikzpicture}%
}

\begin{document}
\bstctlcite{IEEEexample:BSTControl}

\title{Synthetic Electric Vehicle Charging Session Generation Using a Conditional Variational Autoencoder
    \thanks{This publication has emanated from research conducted with the financial support of Taighde Éireann – Research Ireland under Grant numbers 12/RC/2302\_P2 and 22/FFP-A/10455.}
}

\author{
    \IEEEauthorblockN{Graeme Kelly\IEEEauthorrefmark{1}, Emilio J. Palacios-Garcia\IEEEauthorrefmark{1}\IEEEauthorrefmark{2}, and Barry P. Hayes\IEEEauthorrefmark{1}\IEEEauthorrefmark{2}}\IEEEauthorblockA{\IEEEauthorrefmark{1}School of Engineering and Architecture, University College Cork, Cork, Ireland}\IEEEauthorblockA{\IEEEauthorrefmark{2}MaREI Centre, Sustainability Institute, University College Cork, Cork, Ireland\\Emails: 120353736@umail.ucc.ie, epalacios-garcia@ucc.ie, barry.hayes@ucc.ie}
}
\maketitle
\copyrightnotice
\begin{abstract}
    The increasing adoption of electric vehicles (EVs) is expected to place significant additional demand on residential distribution networks, creating a need for realistic charging datasets for planning and simulation studies. However, access to real-world EV charging data is often limited due to privacy constraints, incomplete records, and restricted availability. This paper proposes a conditional variational autoencoder (CVAE) for the generation of synthetic EV charging sessions from real transaction-level charging data. The model is trained on engineered session features describing plug-in duration, charging duration, delivered energy, charging delay, and cyclical time-of-week, while conditioning on day of week and managed charging status. A Gaussian negative log-likelihood (NLL) reconstruction loss is employed to model feature-wise heteroscedastic uncertainty, and the latent space is regularised using a Kullback–Leibler (KL) divergence term. The statistical fidelity of the generated data is evaluated using distributional metrics and downstream task performance through the Train-on-Synthetic-Test-on-Real (TSTR) protocol. Results demonstrate that the proposed approach produces synthetic EV charging sessions that preserve key statistical properties of the original dataset while supporting predictive modelling tasks.
\end{abstract}


\section{Introduction}
Residential energy systems are undergoing rapid electrification due to the increasing adoption of low-carbon technologies such as electric vehicles (EVs), photovoltaic (PV) systems, and heat pumps. Among these technologies, EVs are expected to impose significant additional demand on distribution networks due to their high power requirements and stochastic charging behaviour~\cite{debImpactElectricVehicle2018}. As EV adoption increases, distribution system operators must assess the potential impacts of large-scale EV integration on network capacity, stability, and congestion.

While accurate models of EV charging behaviour are essential for power-system planning and simulation studies, access to real-world charging data is often limited due to privacy concerns, data ownership restrictions and incomplete datasets. These challenges restrict the availability of high-resolution EV charging profiles. In this context, synthetic data generation techniques provide a promising solution by enabling the creation of statistically representative charging profiles without exposing sensitive user information. The generation of such datasets can support a wide range of applications, including network planning, flexibility studies, and scenario analysis~\cite{liCreationValidationLoad2021}.

The modelling of residential electricity demand has been widely studied for power system analysis and planning. Early approaches generated synthetic demand profiles using statistical models such as empirical distributions, autoregressive methods, and clustering techniques \cite{duqueConditionalMultivariateElliptical2021, talbotCorrelatedSyntheticTime2020}. While effective at reproducing aggregate demand characteristics, these methods often fail to capture complex multivariate and temporal dependencies. Generative Adversarial Network (GAN) based approaches have also been explored for energy demand modelling~\cite{huMultiLoadGANGANBasedSynthetic2024,wenRegionalSolarForecasting2023}, although training instability and mode collapse can limit their ability to capture full data variability.

Variational autoencoders (VAEs) provide an alternative generative modelling framework that learns a probabilistic latent representation of the data~\cite{kingmaAutoEncodingVariationalBayes2022}. By optimising a variational lower bound on the data likelihood, VAEs enable stable training while learning compact latent spaces that capture the underlying structure of complex datasets. VAEs have been successfully applied to energy-related problems, including the generation of residential load profiles \cite{panDataDrivenEVLoad2019, razghandiVariationalAutoencoderGenerative2022}. Conditional variants of VAEs (CVAEs) further extend this framework by allowing generation to be conditioned on auxiliary variables, enabling greater control over generated samples~\cite{chaiFaradaySyntheticSmart2024}.

While previous works such as~\cite{chaiFaradaySyntheticSmart2024} have shown the potential of CVAE to generate daily consumption profiles, conditioned on the type of installed low-carbon technologies, this paper focuses on the generation of synthetic individual EV charging sessions that preserve key statistical characteristics of the original dataset. The key contributions are as follows:

\begin{enumerate}
    \item Development of a CVAE for generating individual EV charging sessions that, in contrast with aggregated load profiles, facilitates its use in hosting capacity studies, as sessions can be selectively added to existing demand.
    \item Trained using real, publicly available EV charging session data comprising over 157,000 records.
    \item Potential of conditioning sample generation on the day of the week and charging management capabilities.
    \item Evaluation of the generated synthetic data using distributional similarity metrics and the utility of the model via Train-on-Synthetic-Test-on-Real (TSTR) approach.
\end{enumerate}

\section{Methodology}\label{sec:methods}
\subsection{Dataset and Preprocessing}
This work used real-world electric vehicle charging data, sourced from the Electric Nation programme \cite{electricnationREALWORLDSMARTCHARGING}, i.e. the CrowdCharge and GreenFlux datasets. The combined datasets contained over 157,000 residential charging sessions recorded across 602 unique charging points during a period that spans from early 2017 to the end of 2018. Each record represented a single charging event $(i)$. A subset of features was selected to capture the key temporal and behavioural characteristics of charging sessions, including: timestamps for both plug-in $t_\text{avi}^{(i)}$ and charging start time $t_\text{s}^{(i)}$, session $\Delta t_\text{ses}^{(i)}$ and active charging $\Delta t_\text{cha}^{(i)}$ durations, energy consumption $E^{(i)}$ and a flag indicating whether the session was managed $s^{(i)} \in \{0,1\}$.

Prior to training, several preprocessing steps were applied to ensure data quality and to transform some features into forms more suitable for the model. Records with missing charging-start timestamps $t_\text{s}^{(i)}$ were filtered out, as this feature was critical. This resulted in the loss of ~46k sessions. Additionally, to reduce the influence of rare, uninformative cases, sessions with a plug-in duration $\Delta t_\text{ses}^{(i)} \leq 5$~min or $\Delta t_\text{ses}^{(i)} \geq 10,000$~min were removed. Note that both local and UTC timestamps were available in the data, but it was determined that the local time would better represent user behaviour.

\begin{table}[t]
    \caption{Engineered input features \& conditions}
    \label{tab:processed_features}
    \centering
    \scriptsize
    \renewcommand{\arraystretch}{1.3}
    \begin{tabular}{p{1.6cm} p{4.35cm} p{1cm}}
        \toprule
        \textbf{Feature}               & \textbf{Description}                                     & \textbf{Units} \\
        \midrule
        $\Delta t_\text{ses\_h}^{(i)}$ & Total session duration                                   & h \\
        $\Delta t_\text{cha\_h}^{(i)}$ & Active charging duration                                 & h \\
        $E^{(i)}$                      & Energy consumed during the session                       & kWh \\
        $\Delta t_\text{gap\_h}^{(i)}$ & Time between $t_\text{avi}^{(i)}$ and $t_\text{s}^{(i)}$ & h \\
        $t_\text{avi\_sin}^{(i)}$      & Sine cyclical encoding of $t_\text{avi\_w}^{(i)}$        & $[-1,1]$ \\
        $t_\text{avi\_cos}^{(i)}$      & Cosine cyclical encoding of $t_\text{avi\_w}^{(i)}$      & $[-1,1]$ \\
        \midrule
        $s^{(i)}$                      & Managed session                                          & $\{0,1\}$ \\
        $\textbf{d}_w$                 & Day of the week                                          & $\{0,1\}^7$ \\
        \bottomrule
    \end{tabular}
\end{table}

From these source variables, the features used for training were constructed. Session duration $\Delta t_\text{ses}^{(i)}$ and charging-duration $\Delta t_\text{cha}^{(i)}$ were converted to hours. Then, the feature $\Delta t_\text{gap}^{(i)} = t_\text{s}^{(i)} - t_\text{avi}^{(i)}$ was derived to capture the difference between charging-start $t_\text{s}^{(i)}$ and plug-in time $t_\text{avi}^{(i)}$ in hours.

The day of week was extracted from the plug-in time $t_\text{avi}^{(i)}$ producing an integer value in $[0,6]$ that was encoded into a binary vector as  $\mathbf{d}_\text{w}^{(i)} \in \{0,1\}^7$ (one-hot encoding). In addition, $t_\text{avi}^{(i)}$ was transformed into a relative timestamp in seconds from the start of each week, denoted as $t_\text{avi\_w}^{(i)}$, and its value mapped into a sine $t_\text{avi\_sin}^{(i)}$ and a cosine $t_\text{avi\_cos}^{(i)}$ component as
\begin{align}
    t_\text{avi\_sin} = & \sin\left(2\pi \frac{t_\text{avi\_w}}{3600 \times 24 \times 7}\right) \\
    t_\text{avi\_cos} = & \cos\left(2\pi \frac{t_\text{avi\_w}}{3600 \times 24 \times 7}\right)
\end{align}
This transformation conserved the periodicity of the day of the week. Note that the denominator represents the total seconds in a week, scaling the plug-in time into a circumference in the interval $[0, 2\pi]$. The set of selected features is listed in Table~\ref{tab:processed_features}.

\subsection{Conditional Variational Autoencoder (CVAE)}
\begin{figure}[t!]
    \begin{center}
        \includegraphics[width=0.98\columnwidth]{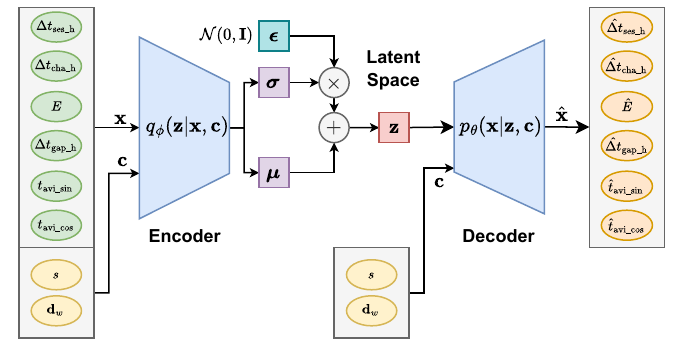}
    \end{center}
    \caption{Architecture of the conditional variational autoencoder}
    \label{fig:cvae}
\end{figure}

A CVAE was adopted to generate synthetic EV charging sessions from the transaction-level features, as shown in Fig.~\ref{fig:cvae}. Given the feature vector $\mathbf{x}\in \mathbb{R}^6$ and the conditional vector $\mathbf{c}\in \mathbb{R}^8$, comprising the one-hot encoding of the day of the week and a binary indicator specifying if a session is managed, the encoder learns an approximate posterior $q_\phi(\mathbf{z} \mid \mathbf{x}, \mathbf{c})$, while the decoder models the conditional likelihood $p_\theta(\mathbf{x} \mid \mathbf{z}, \mathbf{c})$.

The proposed model used a fully connected multilayer perceptron architecture for both the encoder and decoder. In the encoder, the feature and condition vectors are concatenated and passed through two hidden layers with ReLU activations. From the final hidden representation, two linear layers produce the latent mean $\boldsymbol{\mu}$ and latent log-variance $\log \boldsymbol{\sigma}^2$ vectors. The latent samples $\mathbf{z}$ are obtained using the reparameterisation trick
\begin{equation}
    \mathbf{z} = \boldsymbol{\mu} + \boldsymbol{\sigma} \odot \boldsymbol{\epsilon}, \qquad \boldsymbol{\epsilon} \sim \mathcal{N}(0, \mathbf{I}).
\end{equation}
The decoder receives the concatenated latent $\mathbf{z}$ and condition $\mathbf{c}$ vectors, and similarly applies two hidden ReLU layers before producing two outputs: a reconstructed feature mean vector and a feature-wise log-variance vector. This corresponds to modelling the reconstructed charging-session features with a diagonal Gaussian likelihood rather than a deterministic point estimate. To improve numerical stability during training, the decoder log-variance was clamped to a fixed range.

The selected encoder input dimension was 14 (i.e., 6 features and 8 conditions). Additionally, both encoder and decoder used two hidden layers of 256 neurons, and the latent space dimension was set to 64. In this way, the model learned a low-dimensional conditional representation of EV charging behaviour while retaining enough flexibility to capture stochastic variation across features.

The model was trained by minimising a variational objective consisting of a reconstruction term and a Kullback–Leibler (KL) regularisation term,
\begin{equation}
    \begin{aligned}
        \mathcal{L}
        & =
        \mathbb{E}_{q_{\phi}(\mathbf{z} \mid \mathbf{x},\mathbf{c})}
        \left[
            -\log p_{\theta}(\mathbf{x} \mid \mathbf{z},\mathbf{c})
        \right] \\
        & \quad
        + \beta\, D_{\mathrm{KL}}\!\left(q_{\phi}(\mathbf{z} \mid \mathbf{x},\mathbf{c})\,\|\,p(\mathbf{z})\right)
    \end{aligned}
    \label{eq:vae_loss}
\end{equation}
where $p(\mathbf{z})$ is a standard normal prior. Unlike earlier mean-squared-error (MSE) formulations, the reconstruction term used a Gaussian Negative Log-Likelihood (NLL), enabling the decoder to learn feature-wise heteroscedastic uncertainty.

\subsection{Training}
Prior to training, the four non-negative features $\Delta t_\text{ses\_h}^{(i)}$, $\Delta t_\text{cha\_h}^{(i)}$, $E^{(i)}$, and $\Delta t_\text{gap\_h}^{(i)}$ were transformed using $log(1+x)$ in order to reduce skewness and improve the representation of heavy-tailed behaviour. The dataset was then randomly split into training, validation, and test sets using a 70/15/15 partition. Feature standardisation was performed using the mean and standard deviation of the training split only, and these statistics were subsequently applied to the validation and test sets in order to avoid data leakage.

Training was carried out in PyTorch \cite{paszke2019pytorch} using the Adam optimiser with a learning rate of $3 \times 10^{-4}$ and mini-batches of size 256. The model was trained for 60 epochs, with the KL regularisation coefficient fixed at $\beta{}=0.5$. A series of hyperparameter sweeps was used to identify the best model parameters. The reconstruction term was computed as a Gaussian NLL over the six continuous output features, while the KL term regularised the latent posterior.

To improve the balance of feature learning, per-feature weighting was introduced in the reconstruction loss. The weighting vector was applied in the same order as the reconstructed features. These weights were normalised to a unit mean so that the overall loss magnitude remained stable while allowing additional emphasis to be placed on more challenging or more important features. This was found to be useful for addressing under-fitting in selected variables. The final tuning of these weights was also performed using sweeps.

During validation, both total loss and per-feature Gaussian NLL were monitored in order to assess convergence and identify any neglected features. At generation time, synthetic samples were produced by drawing $z \sim \mathcal{N}(0, \mathbf{I})$, fixing the desired condition vector, and decoding to obtain the feature means and variances. Stochastic samples were then obtained from the predicted Gaussian output distribution. This procedure enables the generation of charging sessions under a specific day of the week and managed/unmanaged conditions while preserving variability in the reconstructed features.

\subsection{Evaluation}
Model performance was evaluated using both optimisation-based and data-based measures. Training and validation losses, together with per-feature Gaussian NLL, were monitored across epochs to assess convergence. The statistical fidelity of the generated data was first assessed through histogram comparisons. Subsequently, a series of metrics were calculated across three pairs of distributions, defined as:
\begin{itemize}
    \item \textit{Base}: real distributions of the training and test dataset;
    \item \textit{Train}: synthetic training data (using the condition vector of the real training dataset) and the real training data;
    \item \textit{Test}: synthetic test data (using the condition vector of the real test dataset) and the real test data.
\end{itemize}

The Wasserstein distance (WD) was used to measure feature-wise similarity, while the sliced WD (SWD) was employed to evaluate distance between joint distributions as it approximates the multivariable WD by averaging the WD across multiple random projections. Preservation of inter-feature relationships was also evaluated using the Spearman correlation mean absolute error (MAE) between distribution pairs. All metrics were computed on normalised datasets to allow for feature-wise comparison and across various sampling rounds due to the stochastic nature of the generative process.

Finally, to assess the model utility, the TSTR evaluation method was applied. In this approach, a predictive model, in this case a Random Forest (RF) estimator composed of 200 decision trees, was first trained on the synthetic dataset and then evaluated on the real test data. The resulting prediction error is then compared with the error obtained when the same model is trained and tested on real data (TRTR). The ratio between the two errors is computed as
\begin{equation}
    \text{TSTR/TRTR Ratio} = \frac{\text{MAE}_{\text{TSTR}}}{\text{MAE}_{\text{TRTR}}}
\end{equation}

Ratios close to 1 indicate similar predictive performance for models trained on synthetic and real data, suggesting that the synthetic dataset preserves the relevant feature relationships required for downstream modelling tasks.

\section{Results}
This section discusses the CVAE performance in generating synthetic EV charging sessions. The evaluation considered model training behaviour, distributional similarity between real and synthetic data, and the model's utility for downstream modelling tasks. All code and data are available in~\cite{EV-CVAE}.

\subsection{Training Behaviour}
\begin{figure}[t]
    \centering
    \includegraphics[width=1.0\linewidth]{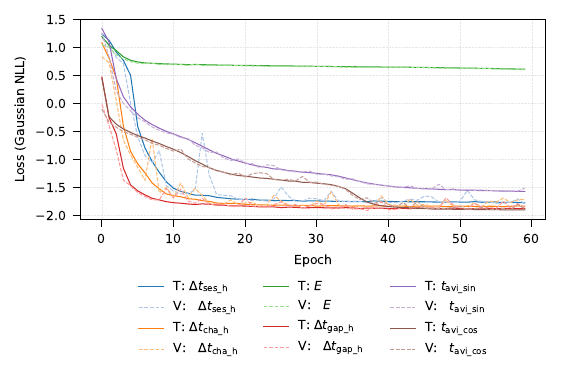}
    \caption{Training (T) and Validation (V) loss per feature.}
    \label{fig:PerFeatureNLL}
\end{figure}
Fig.~\ref{fig:PerFeatureNLL} shows the training and validation reconstruction loss for each modelled feature throughout the training process. The loss curves for most variables decrease rapidly in the early training epochs and gradually stabilise after approximately 40--50 epochs, indicating model convergence to a stable solution.

The similarity between the training and validation loss trajectories suggests that the model does not exhibit significant overfitting. Features related to charging duration and time gaps converge to relatively low loss values, indicating that the latent representation effectively captured these variables. In contrast, variables associated with energy consumption and cyclical temporal representations exhibit slightly higher final loss values, reflecting their greater variability and complexity.

\subsection{Distribution Similarity}
\begin{figure}[!t]
    \centering
    \includegraphics[width=0.94\linewidth]{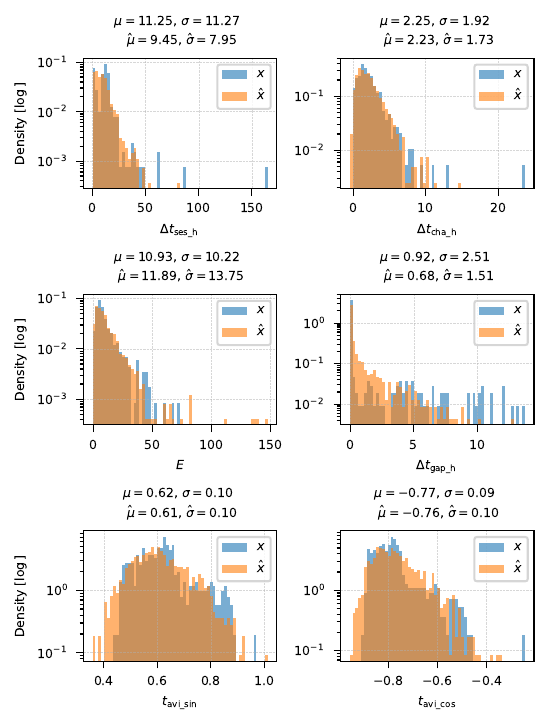}
    \caption{Density-normalised histograms comparing feature distributions between real (n = 470) and synthetic samples (n = 1000) in the unscaled space.}
    \label{fig:Histograms}
\end{figure}

Fig.~\ref{fig:Histograms} compares the marginal feature distributions of the synthetic samples $\hat{x}$ with the real test data $x$, conditioned to be an unmanaged Wednesday ($s=0$, $d_\text{w}=2$). Each subplot represents an individual feature denoted by the X-Axis labels. The mean and standard deviation are given for real ($\mu$, $\sigma$) and generated ($\hat\mu$, $\hat\sigma$) data on top. The Y-axis is shown on a logarithmic scale. The results demonstrate that the synthetic data reproduces the overall shape of the real distributions.

For energy consumption $E$ and charging session duration $\Delta t_\text{cha\_h}$, the generated samples captured the strong skew toward shorter sessions observed in the real data, with probability densities that closely match the empirical distribution. The synthetic energy consumption also followed the general structure of the real distribution. However, a few extreme values are visible in the upper tail of $E$, reflecting the existence of sporadic but intense EV charging sessions.

The temporal features $t_\text{avi\_sin}$ and $t_\text{avi\_cos}$ were also broadly preserved, with minor deviations in the extreme regions of the distributions. In contrast, the session duration $\Delta t_\text{ses\_h}$ exhibited a small bimodal structure in the real data that appeared slightly smoothed by the generative model. However, this is consistent with the VAE's regularisation behaviour. Additionally, the charging start gap $\Delta t_\text{gap\_h}$ showed an irregular and highly skewed distribution in the real dataset, with a dominant peak at small values followed by several smaller spikes, a structure that the CVAE model did not fully capture.

\begin{figure}[t!]
    \begin{center}
        \includegraphics[width=0.98\columnwidth]{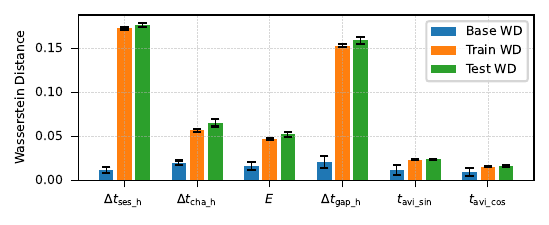}
    \end{center}
    \caption{Per-feature \textit{Base}, \textit{Train}, and \textit{Test} WD. Average of 10 rounds.}
    \label{fig:wd}
\end{figure}

The feature-wise WD in Fig. \ref{fig:wd} support the graphical insights from the histograms. The \textit{Base} WD (blue) is calculated between the real training and test datasets and represents the lowest WD bound. The \textit{Train} WD (orange) and \textit{Test} WD (green) measure the distance between the synthetic and real training, and synthetic and real test datasets, respectively. The feature-wise \textit{Test} WD remains close to the \textit{Train} WD, indicating the model did not overfit the training dataset. In terms of individual features, $\Delta t_\text{ses\_h}$ and $\Delta t_\text{gap\_h}$ showed the highest WD (lowest similarity) followed by $\Delta t_\text{cha\_h}$ and $E$, and finally $t_\text{avi\_sin}$ and $t_\text{avi\_cos}$ with the lowest WD.

\begin{table}[b]
    \centering
    \caption{Global normalised similarity metrics comparison over 10 rounds of samples. Standard deviation in brackets.}
    \label{tab:global_metrics}
    \begin{tabular}{@{}lrr}
        \toprule
        \textbf{Data} & \textbf{SWD}  & \textbf{Spearman-MAE}\\
        \midrule
        Base          & 0.017 (0.001) & 0.007 (0.001) \\
        Train         & 0.050 (0.001) & 0.014 (0.001) \\
        Test          & 0.058 (0.002) & 0.015 (0.001) \\
        \bottomrule
    \end{tabular}
\end{table}

Table~\ref{tab:global_metrics} presents the normalised SWD and Spearman correlation MAE (Spearman-MAE) that evaluate the overall model performance. The \textit{Base} SWD and Spearman-MAE are shown again as lower bounds. The \textit{Train} and \textit{Test} SWD and Spearman-MAE values remain close, with slightly worse performance on the test dataset, but not significant enough to suggest overfitting. The \textit{Test} SWD of 0.058 indicates a good level of agreement between the joint feature distributions, while the Spearman-MAE of 0.015 implies that rank-based relationships were largely preserved. These metrics suggest that the generated samples captured much of the original distribution structure and maintained many of the monotonic dependencies present in the real charging data.

\subsection{Model Utility Evaluation}
Two independent RF regressor models were fitted: one on a synthetic training dataset and the other on the real training dataset, both incorporating the condition vector as an input feature. The $\text{MAE}_\text{TSTR}$ and $\text{MAE}_\text{TRTR}$ were then calculated on an unseen real test dataset where the target feature was removed. Table~\ref{tab:tstr_features} reports their values and their ratios across multiple target features. The charging duration $t_\text{cha\_h}$ exhibited the closest agreement (1.075), suggesting that this feature was well represented in the synthetic dataset. The consumed energy $E$ and the start gap $t_\text{gap\_h}$ showed slightly higher ratios of 1.101 and 1.114, respectively, indicating modest performance degradation when training on synthetic data. The session duration $t_\text{ses\_h}$ presented the largest ratio (1.207), suggesting this feature was more difficult to reproduce accurately.

Table~\ref{tab:tstr_kwh_scenarios} further evaluates the RF models' performance for predicting consumed energy $E$ across different days of the week $d_\text{w}$ and managed charging scenarios $s$. Note that RF models are not retrained at this stage. Across all cases, the TSTR/TRTR ratios remained above but close to 1, indicating that the models trained on synthetic data achieved prediction errors slightly higher than those trained directly on the real dataset. Slightly higher ratios were also observed in the managed charging scenarios, suggesting that the additional variability introduced by managed charging behaviour may be more challenging for the generative model to capture fully. These metrics suggest that the synthetic data preserves marginal distributions and key structural relationships of the original dataset, supporting its suitability for downstream energy modelling tasks.

\begin{table}[t]
    \centering
    \caption{TSTR/TRTR ratios across target features for the representative evaluation scenario on Test set.}
    \label{tab:tstr_features}
    \begin{tabular}{lrrr}
        \toprule
        \textbf{Target Feature}  & $\textbf{MAE}_\textbf{TSTR}$ & $\textbf{MAE}_\textbf{TRTR}$ & \textbf{Ratio} \\
        \midrule
        $E$                      & 0.425                        & 0.338                        & 1.101 \\
        $\Delta t_\text{ses\_h}$ & 0.612                        & 0.507                        & 1.207 \\
        $\Delta t_\text{gap\_h}$ & 0.666                        & 0.598                        & 1.114 \\
        $\Delta t_\text{cha\_h}$ & 0.394                        & 0.336                        & 1.075 \\
        \bottomrule
    \end{tabular}
\end{table}

\section{Conclusion \& Future Work}
This paper has presented a CVAE for generating synthetic EV charging sessions from transaction-level data. The proposed model was able to learn a probabilistic latent representation of key session attributes, including session and charging duration, delivered energy, and temporal features, while conditioning on contextual variables such as day of week and managed charging status. By modelling feature-wise uncertainty through a Gaussian reconstruction loss, the approach captured the joint statistical structure of charging sessions and enabled the generation of realistic synthetic data.

The fidelity of the generated data was evaluated using statistical similarity metrics and downstream task performance through the TSTR method. The results evidenced that the synthetic data preserves many of the statistical properties of the original dataset while maintaining predictive utility for modelling tasks. These findings suggest that CVAE models can provide a practical approach for producing synthetic EV charging datasets suitable for energy system analysis.

\begin{table}[t]
    \centering
    \caption{TSTR/TRTR ratios for $E$ prediction on Test set across day of week $\mathbf{d}_\text{w}$ and $s=0$ (unmanaged) / $s=1$ (managed)  scenarios}
    \label{tab:tstr_kwh_scenarios}
    \setlength{\tabcolsep}{3.5pt}
    \begin{tabular}{@{}lccccccc@{}}
        \toprule
                                & \multicolumn{7}{c}{\textbf{Day of week}}\\
        \textbf{Managed Status} & Mon   & Tue   & Wed   & Thu   & Fri   & Sat   & Sun \\
        \midrule
        Unmanaged               & 1.076 & 1.021 & 1.025 & 1.084 & 1.043 & 1.083 & 1.022 \\
        Managed                 & 1.158 & 1.093 & 1.097 & 1.167 & 1.109 & 1.079 & 1.088 \\
        \bottomrule
    \end{tabular}
\end{table}
Future work will focus on improving the representation of rare events and heavy-tail behaviours, as they were not fully captured by the learned latent representation. Potential directions include alternative conditioning strategies, adjustments to regularisation and weighting within the training objective, as well as extensions to incorporate higher-resolution temporal charging profiles. Such developments could further improve the realism and utility of synthetic charging datasets. In addition, the integration of the generated synthetic EV charging sessions into a low-voltage distribution network simulation framework to evaluate hosting capacity under large-scale EV demand forms part of our current research efforts.

\bibliographystyle{IEEEtran}
\bibliography{references}

\end{document}